# Emotion in Good Luck and Bad Luck: Predictions from Simplicity Theory

**Jean-Louis Dessalles (dessalles@telecom-paristech.fr)**
Telecom PARISTECH, 46 rue Barrault,
F-75013 Paris, France
&
ILCAA, Tokyo University of Foreign Studies,
Fuchu-shi, 183-8534, Tokyo, Japan

### Abstract

The feeling of good or bad luck occurs whenever there is an emotion contrast between an event and an easily accessible counterfactual alternative. This study suggests that *cognitive simplicity* plays a key role in the human ability to experience good and bad luck after the occurrence of an event.

## Good Luck and Bad Luck

Situations spontaneously associated with good luck or bad luck are an important source of emotion. They are frequent in daily life: missing (or catching) the train by five seconds, forgetting one's cell phone the very day one is late for an important appointment, finding a banknote on the ground, etc. They are heavily used in popular fiction, precisely to arouse emotion: the gun gets jammed just at the right (or bad) time, the heroine defuses the bomb just before it explodes, etc. Regarding oneself or someone else as lucky or unlucky on specific occasions may induce gratitude or guilt, and for those who downplay the role of chance, intense feelings of good or bad luck may strengthen supernatural beliefs (Teigen & Jensen, *in press*). Reasoning about good luck and bad luck may also significantly influence rational judgment (Roese, 1997; Wohl & Enzle, 2003).

The feeling of having good or bad luck is a clear-cut phenomenon. Different individuals have consistent views of which situations can be regarded as bad or good luck (what the present study will confirm). This ability therefore gives rise to a well-posed problem, worth investigating. Previous studies have identified various parameters that control the feeling of luck. These include physical or temporal closeness (Kahneman & Varey, 1990; Teigen, 1996; Roese, 1997; Pritchard & Smith, 2004), deviation from norms and expectations (Kahneman & Miller, 1986), mutability of causes (Kahneman & Miller, 1986; Byrne 2002, 2007) and controllability (Roese, 1997).

Many authors have acknowledged the prime importance of counterfactuals in any situation that generates a strong feeling of good or bad luck. Individuals systematically go through thoughts such as "If only…" or "I almost…" when regarding situations as (un)lucky. The theoretical treatment of counterfactuals in general, and in emotional situations in particular, remains however complex, as a multitude of determining factors seem to be involved.

The purpose of the present study is to propose a new perspective on the phenomenon, imported from two other scientific domains. One is the study of narrative relevance (Dessalles 2008a). Spontaneous conversations are replete with stories about (un)lucky episodes, and the laws of interestingness seem to apply to them. The other import is the mathematical notion of complexity, which is involved in several important cognitive phenomena (Chater 1999; Chater & Vitányi, 2001).

After mentioning existing attempts to capture the good/bad luck phenomenon formally, I will briefly present the Simplicity Theory and its first predictions concerning our problem. I will then present a study that seems to corroborate those predictions. Then, I will consider situations in which individuals adopt causal thinking. The results and the scope of the theory will be discussed in a last section.

## Formal accounts of luck

Various determining factors have been identified that control the intensity of luck. One of them is the low probability of the (un)lucky event $s$. According to Rescher (1995:211), the intensity of luck is given by $L = E(1–p)$, where $E$ measures the difference that the occurrence of $s$ makes for the interests at stake, and $p$ is its probability. This formula has two major drawbacks. First, contrary to intuition, it does not distinguish moderately unlikely outcomes from highly unlikely ones, as both would provide emotion roughly equal to $\Delta u$. Second, as pointed out by Teigen (2005), it fails to capture the crucial presence of a counterfactual. As shown by Teigen in various studies, the amplitude of (un)luck is controlled by the 'distance' to an alternative outcome that would have provided an emotional contrast. Teigen (2005) represents these effects through the formula: $L = \Delta u / D$, where $\Delta u$ is the difference in 'utility' between the counterfactual $s_2$ and the actual situation $s_1$, whereas $D$ represents the 'distance' between $s_1$ and $s_2$.

This formula makes predictions that are much closer to observation, and thus represents a significant progress in comparison with Rescher's initial proposal. It has, however, its limitations. First, the influence of low probability, as identified by Rescher, is lost. The problem is illustrated in figure 1, where the feeling of unluck after missing the winning sector (in color) in a wheel of fortune game is stronger in (b) than in (a). Second, the notion of 'utility', imported from economics, does not account for situations of pure surprise ('I almost got six on all dice'). Third, the notion of distance is not properly defined. Sitting next to a lottery winner doesn't make you feel unlucky; you might however feel unlucky to have played her winning numbers, but a week too soon. Lastly, Teigen's formula fails to capture one property of counterfactual $s_2$ that contributes to (un)luck, namely its simplicity. In figure 1(c), the winning sectors (in color) of the wheel of fortune occupy the same area as in (a) and the distance to the landing site is the same in both cases. Judgment of bad luck is, however, stronger in (a) than in (c). This phenomenon, due to the greater complexity of the counterfactual in (c), is not predicted by Teigen's formula.

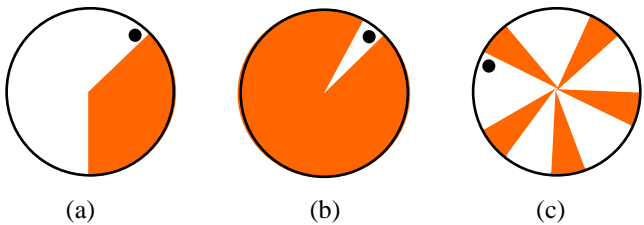

(a)         (b)         (c)

Figure 1: Three examples of near miss

We will propose an alternative account, based on Simplicity Theory. It can be formulated in an informal way:

*(Un)lucky events are situations that occurred despite of simple, easily accessible alternatives.*

## Simplicity Theory

Simplicity Theory (ST) (formerly called 'Complexity Drop Theory') has been developed to predict how people select events worth to tell. It has applications in the study of spontaneous conversations, of narratives, of news, and in the definition of subjective probability (Dessalles, 2006; 2008a). ST's main principle can be stated:

*Interesting situations are those which are 'too' simple.*

ST uses the notion of cognitive complexity, which is a slightly modified version of the mathematical notion know as Kolmogorov complexity.

*The complexity $C(s)$ of a situation s is the size of the ideal minimal description of s that is available to the observer.*

(the last restriction is crucial for the notion to be useful in cognitive science). The concept is much less trivial than it seems at first sight, and has given rise to a growing literature since its definition in the years 1960.

ST uses two notions of complexity. The second notion is *generation* complexity.

*$C_w(s)$ is the minimal size of the parameters to be set for the 'world' w to generate situation s.*

To compute $C_w(s)$ of a lottery draw, for instance, one adds up the descriptions of all drawn numbers, as the 'world' (in this case, the lottery machine) had to 'choose' them independently. Note that the notion refers, not to any objective world, but to the observer's perception of the world. ST's central notion is unexpectedness, noted $U(s)$.

$$U(s) = C_w(s) - C(s) \quad (1)$$

A situation is unexpected if it is 'too' simple, *i.e.* simpler to describe than to generate. In the lottery example, a 'remarkable' lottery draw such as 22-23-24-25-26-27 is unexpected, since $C$ is much smaller than $C_w$. It only requires to instantiate 22 and to mention that it is a continuous series. $C$ thus spares five instantiations by comparison with $C_w$. Hence a strong feeling of unexpectedness if such a draw actually occurs (Dessalles 2006). This definition of unexpectedness accounts for various cognitive abilities, such as the perception of coincidences (Dessalles, 2008b) and of interestingness (Dessalles, 2008a; Dimulescu & Dessalles, 2009) (see details on www.simplicitytheory.org). It is consistent with the observation that 'contrast' (what we call unexpectedness) is more relevant than (standard) probability to explain surprise (Teigen & Keren, 2003).

Complexity is usually linked to probability $p_0$ thanks to the following formula $p_0 = 2^{-C_{w_0}}$, where $w_0$ is a blank world (Solomonoff, 1978). This formula is, however, unsatisfactory, as it assigns a virtually zero probability to most situations of daily life, as they depend on a huge quantity of parameters. If we replace the blank world $w_0$ by the observer's model $w$ of the actual 'world', we get $p_w = 2^{-C_w}$, which corresponds to the usual definition of 'objective' probability. In a lottery, for instance, $p_w$ is the same for all draws. ST (Dessalles 2006) defines *subjective* probability $p$ by subtracting cognitive complexity $C$ from $C_w$. We get:

$$p = 2^{-U} \quad (2)$$

Hence the statement about unexpected events being 'too' simple. In ST's framework, the concept of probability is a derived notion and should be replaced by the notion of unexpectedness to account for many aspects of cognition.

To account for good luck and bad luck, we must say how emotion is related to simplicity (Dessalles, 2008a). Let's call $E(s)$ the (always positive) intensity of the emotional experience caused by situation $s$.

$$E(s) = E_h(s) + U(s) \quad (3)$$

$E_h(s)$ is the hypothetical emotional intensity attached to the occurrence of $s$. It corresponds to a not unexpected experience (when $U = 0$). In many cases, $E_h(s) = V(s)$, where $V$ is a utility function. Events that were complex for the world to produce ($C_w$ large) arouse more intense emotion when they occur, as they are more unexpected. Using (2), (3) can be rewritten: $e(s) = e_h(s)/p(s)$, where $e_h$ and $e$ stand for non-logarithmic emotions. The cognitive complexity $C(s)$ decreases $E(s)$ in (3). It acts like an emotional 'tax' paid for considering the event.[1]

---

[1] In (2), $U$ must remain positive. In (3), $U$ may be negative, but $E$ must be positive. These constraints can be used to define the *relevance* of events (Dessalles, 2008a).

If $s$ is not an event, but an anticipated situation, the expected emotion can be expressed using utility function $V$:

$$E_h(s) = V(s) - U(s) \qquad (4)$$

The perspective of a situation that is complex for the world to produce ($C_w$ large) arouses less emotion. In the non-logarithmic domain, equation (4) reads $e_h(s) = v(s) \times p(s)$.

In causal reasoning, we suppose that expected emotion propagates through causal links (Dessalles 2008). If a known emotional situation $s$ is believed to result from situation $s'$, then $E_h(s) = E_h(s')$. Using conditional complexity, we may write:

$$U(s) = U(s') + C_w(s|s') \qquad (5)$$

By adding $E_h(s)$ to both sides, we get:

$$E(s') = E(s) - C_w(s|s') \qquad (6)$$

## ST's Predictions

In the absence of any precise counterfactual, as when one's house is struck by lightning, (3) provides a definition of luck, in line with (Rescher 1995):

$$L_1 = E_h(s_1) + U(s_1) \qquad (7)$$

To assess the expected emotion $E_h(s_1)$ in such case, individuals may recall a known situation $s$ of lightning on a house (or imagine it), and consider $E_h(s_1) = E_h(s) = L_1(s) - U(s)$.

In wheel of fortune situations, the expected emotional intensity $E_h(s_+)$ of winning corresponds to landing on a winning site $s_+$. The colored segment in figure 2 represents the winning sector in a linear version of the wheel of fortune. The complexity of landing on $s_+$ is $C_w(s_+) = \log_2 l_0$. This is the number of bits required by the 'world' to choose a landing position. According to (4), the maximum value of $E_h(s_+)$ is obtained for typical, i.e. maximally complex $s_+$: $C(s_+) = \log_2 l_2$. This is the number of bits required to discriminate among all winning positions. We get:

$$\max E_h(s_+) = V(s_+) - \log_2 l_0/l_2$$

This corresponds to the classical expected utility in the non-logarithmic domain.

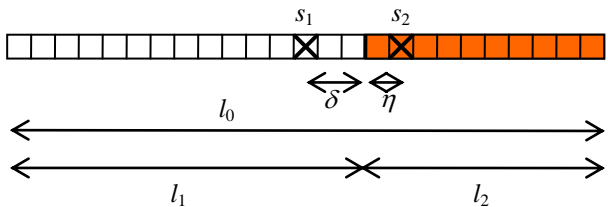

Figure 2: Discrete bounded near miss

When playing with a wheel of fortune, individuals acknowledge that the probability of landing in various sectors of the roulette is constant, but that landing close to a winning sector involves more intense bad luck (Teigen, 1996). Let us considered a linear version of the problem (figure 2).

After the draw, possibly for (self-)narrative purposes, individuals pick the situation $s$ that maximizes emotional intensity $E(s)$. It may be the actual situation $s_1$, as in (7), or a counterfactual one $s_2$. Individuals are supposed to opt for the computation that gives the more intense emotion. In the counterfactual case, $s_1$ is seen as an intermediary step toward $s_2$. (3) and (6) give a new value for $E(s_1)$: $E_c(s_1) = E_h(s_2) + U(s_2) - C_{wc}(s_2|s_1)$. Luck is measured by the emotional gap between both emotions for $s_1$:

$$L_2 = E_h(s_2/s_1) + U(s_2) - C_{wc}(s_2|s_1) \qquad (8)$$

Conditional $E_h(s_2/s_1)$ means that the expected emotional intensity is assessed using the actual emotional intensity of $s_1$ as baseline. The counterfactual nature of $s_2$ requires the introduction of a fictitious world $wc$ that is able to keep a memory of $s_1$ to generate $s_2$. The term $C_{wc}(s_2|s_1)$ is the minimal price to pay for the 'If…'. It represents the size of the minimal parameter modifications that the observer can imagine for the 'world' to have generated $s_2$ instead of $s_1$.

In the case of figure 2, $E_h(s_2/s_1) = V(s_+)$, and $C_{wc}(s_2|s_1) = 1 + \log_2(\delta+\eta)$, which is the amount in bits needed to indicate the (non zero) targeting shift to the right toward $s_2$. On the other hand, $C_w(s_2) = \log_2 l_0$ and $C(s_2) = 1 + \log_2(1+\eta)$ (one bit to choose the left edge of the winning region, plus the representation in bits of the (possibly null) shift to reach $s_2$). We get: $L_2 = V(s_+) + \log_2 l_0 - \log_2(\delta+\eta)(1+\eta) - 2$. Taking $\eta = 0$ to maximize the intensity of unluck:

$$L_2 = V(s_+) + \log_2 l_0/\delta - 2 \qquad (9)$$

The experience of bad luck in this near miss experience is an increasing function of the missed opportunity $V(s_+)$ and of the number $l_0$ of possibilities, and a decreasing function of the miss $\delta$.

If the counterfactual is assessed against the expected emotion, here max $E_h(s_+)$, instead of $s_1$, we get:

$$L_3 = V(s_+) + \log_2 l_2/\delta - 2 \qquad (10)$$

This model accounts for the fact that when $s_2$ is more complex, as in figure 1(c), the intensity of (un)luck is smaller. We have $C(s_2) = \log_2 k + 1 + \log_2(1+\eta)$, where $k$ is the number of winning regions. The intensity of luck is thus diminished by $\log_2 k$.

The extension to the continuous case is straightforward (figure 3). We suppose that the space is bounded to the left but not to the right. If we call $\alpha$ the landing precision, then $C_w(s_2) = \log_2(l_0/\alpha)$, as we need that number of bits to decide where to stop.[2] As previously, $C_{wc}(s_2|s_1) = \log_2(\delta+\eta)/\alpha$, and $C(s_2) = \log_2(1+\eta/\alpha)$. After taking the best choice $\eta = 0$, we get:

$$L_2 = V(s_+) + \log_2 l_0/\delta - 1 \qquad (11)$$

---

[2] This supposes that there is a way to delimit numbers in the algorithm.

(the one-bit difference with (9) comes from the fact that the winning region has only one edge). Equation (11) accounts for emotions described by the expression: "fall short of the goal".

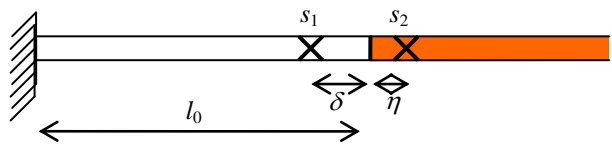

Figure 3: Continuous unbounded near miss

Equations (9) and (11) define the intensity of luck, but not only. They predict what the counterfactual situation $s_2$ will be (what many models of counterfactual thinking omit to do). Individuals pick the alternative $s_2$ that realizes the best compromise between high emotion $E(s_2)$ and low counterfactual complexity $C_{wc}(s_2/s_1)$.

## Nine Stories

The following experiment was conducted to validate the predictions of the model. We tested 61 participants who accepted to pass the test on a Web site (www.dessalles.fr/expe/histoires). All contacted individuals had a high level of academic education, though in domains different from psychology or language sciences (mainly students in engineering). Nine short stories were presented to them (see Table 1). Each involved two or three choices. Instructions invited participants to choose options that made emotion maximal. Some choices irrelevant to the present study (such as the age of the victim in story S9) have not been exploited. Answers given after less than 20 sec. of reading were automatically discarded (median answering time per story was 90 sec.), which leaves us with a minimum of 56 answers per story. Presentation order (stories and options) was randomized.

Table 1: Abridged translation of the stories
(originals on www.dessalles.fr/expe/histoires).

S1- René is a railway worker. He works at the border, at a place where signals must me manually transmitted between the two networks. There is single-track line at [**9** (71*) / **23** (21) / **15** (7)] km from René's post. That day, René forgot to send the signal as a train crossed the border. He eventually did, but [**ten** (59*) / **fifty** (21) / **thirty** (20)] seconds before that, another train had entered the single-track line. The collision killed one of the two drivers.

S2- Lucas was heading for the metro station. At [**30** (71*) / **100** (20) / **800** (9)] m from the station, he stopped to lace up his shoe. As he arrived on the platform, the doors of the train closed in front of him. He had to wait [**25** (89*) / **15** (9) / **6** (2)] minutes for the next train.

S3- Michèle has been playing lotto every week for [**6** (84*) / **4** (11) / **2** (5)] years. On December [**19** (70*) / **3** (18) / **12** (12)], she told [**two** (60*) / **four** (32) / **three** (9)] friends of hers that she would stop playing. They persuaded her to bet for the special Christmas draw, on December 26. She did and won 62 000 Euros.

S4- Jacques was badly injured at his workplace by a defective machine on November 7. The defect had been previously notified and the machine was planned to be repaired on [**November** 8 (75*) / **November 17** (12) / **December 18** (12)].

S5- Florence works in a biology lab. Her two-[**year** (84*) / **month** (11) / **week** (5)] experiment on cell cultures was ruined by a student who knocked over a shelf. This broke [**all boxes containing** (35) / **a bottle of formalin that fell on** (45) / **the automatic device nourishing** (20)] the cell cultures. Florence was furious. She discovered that the student was the son of [**her neighbor** (67*) / **her former PE teacher** (15) / **the piano teacher of her sister** (18)].

S6- A young writer is admitted to Magalie's emergency department at the hospital. Her condition deteriorates. [**8** (66*) / **4** (21) / **6** (14)] infectious agents may explain the illness. Magalie sends samples to the lab and tests are conducted in parallel. It takes [**seven** (79*) / **three** (16) / **five** (5)] hours to get the result and the patient is saved at the last minute. Magalie remembers that she saw the name of the virus in [**the media, as well-know singer recently died of it** (52*) / **the record of another patient** (28) / **a specialized journal** (21)].

S7- For [**four months** (76*) / **two months** (21) / **two weeks** (3)] I was thinking of changing my cell phone. I eventually went to SFR Thursday at 1pm. I had to pay part of it because I was lacking 1000 points. [**Thursday** (74*) / **Friday** (21) / **Tuesday** (5)] evening, I received an offer: "change your phone, SFR offers you [**1500** (55) / **4000** (38) / **500** (7*)] points".

S8- Ms Tsuda's daughter had invited [**two friends** (71*) / **all girls in her class** (17) / **four friends** (12)] to her house. One of them was late. She had left her own house long ago. Ms Tsuda walked toward the girl's house and arrived at a level crossing, located at [**200** (55*) / **500** (24) / **900** (21)] m from Ms Tsuda's house. There was indeed an accident involving a young girl. It turned out that the invited girl was not involved and was late because of a detour caused by the accident.

S9- Helen, retired teacher, fainted as she was walking in the woods. She was found by [**a retired couple** (49) / **a colleague teacher** (26) / **a member of her bridge club** (25)] who called the rescue team. Helen would not have survived if she had reached the hospital [**half an hour** (77*) / **one hour** (16) / **one hour and a half** (7)] later.

Note: Choices irrelevant to the present study are not shown here. Numbers in parentheses indicate percentages. Asterisk indicates significance (p < 0.001). Underlined numbers indicate model predictions.

As shown in Table 1, most results were significant and 19 of the 21 majority choices are congruent with the model's predictions.

## Analysis

Some results are commented now in the light of the theory.

*Emotions*: The intensity of the actual event, $E(s_1)$, was tested in story S2 (Lucas's waiting time), and in story S5 (duration of Florence's lost experiment). Unsurprisingly, majority choices make $E(s_1)$ maximal. In story S7, the third choice (number of points offered) influences $E_h(s_2)$: option "500", which would lead to a smaller value of $E_h(s_2)$, was discarded by participants.

*Counterfactual simplicity*: In story S8, counterfactual $s_2$ corresponds to the invited girl (G) being involved in the accident ('it could have been her'). Both majority choices in S8 tend to make $s_2$ simpler, in agreement with equation (8). Participants clearly preferred that the invited girl (G) be one among 2 (71%) instead of one among 5 (17%) or 30 (12%), thus making the minimal description of G smaller by $\log_2 n - 1$ in comparison with $n = 5$ and $n = 30$. Similarly, by choosing the closest location (200m (55%)) instead of 500m or 900m for the counterfactual accident, they saved bits on $C(s_2)$ ($\log_2(500/200)$ and $\log_2(900/200)$).

*Duration before near miss*: In story S7, participants judged important that the hero hesitated four months (76%) instead of two months or two weeks before buying her/his telephone. We are in a case of unbounded near miss, and as predicted by equation (11), participants preferred the largest value for $L$. The same phenomenon explains the strong preference for the fact that Michèle has been playing for 6 years (84%) in story S3 (in this case, the winning 'sector' is $s_1$ and it is reached, but the computation is identical).

*Proximity in near miss*: Equation (11) predicts that emotion is maximum when one ends up close to the border between 'winning' and 'loosing' sectors ($\delta$ small). Several stories represent near miss situations. In S1, the train accident would have been prevented if the signal had been sent $k \times 10$sec before ($k = 1$ preferred (59%)); In S4, the worker would not have been badly injured if the accident had occurred $k$ days later ($k = 1$ preferred (75%)); in S7, the cost would have been saved if the purchase had been made $k$ days later ($k = 1$ preferred (74%)); in S9, Helen would have died if her admission had been delayed by $k \times 30$min ($k = 1$ preferred (77%)).

## Causal Thinking in Good or Bad Luck

When confronted with events they perceive as (un)lucky, people tend to construct causal explanations for why these events happened (Pritchard & Smith, 2004). Causal thinking may produce counterfactuals by negating causes of the actual event, but also by enabling conditions for the counterfactual (Byrne, 2007). In what follows, we show how causal thinking can be accounted for within the ST framework.

Suppose that a cause $s_3$ can be found to explain $s_1$. If we use (5) together with (7), we get:

$$L_1 = E_h(s_1) + U(s_3) + C_w(s_1|s_3) \quad (12)$$

This relation shows that unexpected causes ($U(s_3)$ large) and materially complex causal links will tend to increase the feeling of (un)luck in the non-counterfactual case.

If $s_4$ is a counterfactual alternative to $s_3$ that would have led to $s_2$, we can compute $L_2$ from $s_3$. Using (8):

$$L_2 = E_h(s_2/s_3) + U(s_2) - C_{wc}(s_2|s_3)$$

We may decompose $C_{wc}(s_2|s_3)$:

$$L_2 = E_h(s_2/s_3) + U(s_2) - C_w(s_2|s_4) - C_{wc}(s_4|s_3) \quad (13)$$

The term $C_{wc}(s_4|s_3)$ measures the mutability of $s_3$ (Byrne, 2007). Equation (13) can be used to find a cause that people will be likely to select as mutable. Let us check these predictions against the experimental results.

*Cause simplicity*: Relation (12) predicts that simple causes ($C(s_3)$ small) will augment emotion since they are more unexpected. This is verified in story S5, where participants preferred the student responsible for the damage to be a neighbor's son (67%) instead of more complex individuals. In story S6, they preferred the virus to have been mentioned in the media (52%), rather than in a medical journal or a medical record where it would have been more complex to discriminate. Story S9 was also designed to test causal simplicity. We expected participants to reject option 'a retired couple', as these individuals would be more complex to discriminate than in the two other options ('a colleague teacher' and 'a member of her bridge club'). However, participants did not show the expected preference (49% *vs.* 26%+25%).

*Causal link complexity*: Relation (12) predicts that materially complex causal links ($C_w(s_1|s_3)$ large) are more unexpected and thus will augment emotion. Story S6 has been designed to check this point. Participants did prefer Magalie's eventual success to go through a seven hour (79%) test to decide between 8 (66%) infectious agents, rather than easier alternatives.

*Causal link simplicity*: Relation (13) conversely predicts that in counterfactual thinking, simple causal links will be preferred ($C_w(s_2|s_4)$ small). In story S1, participants chose the shortest distance between the railway worker's faulty action and its effect (71%); in story S2, they preferred Lucas to lace up his shoe close to the station (71%). In both cases, the material simplicity of the causal link diminishes the counterfactual complexity from the cause ('if he had sent the signal…', 'if Lucas had not paused to lace his shoe…') to the counterfactual effect. We failed to show the same effect in story S5, where we expected participants to chose the simpler causal mechanism ('broke all the boxes') instead of more complex ones ('broke a bottle of formalin'; 'broke the nourishing device'). The probable reason is that a simple causal link is preferable if one adopts Florence's counterfactual thinking, whereas a complex causal link is preferable if we only consider the newsworthiness of the story, what some participants seem to have done despite the instructions.

## Discussion

The strong point of this study was to show the relevance of the notion of complexity in the study of the perception of luck. Many judgments about (un)lucky situations are not explained by variations of probability (even perceived probability) (Teigen, 1996). However, they vary in a systematic way according to variations in complexity. We tried to connect people's attitude toward luck with predictions from Simplicity Theory, with some positive results.

Another positive aspect of the study is to highlight several intervening factors that have gone unnoticed in previous studies, such as the simplicity of the counterfactual situation (story S8), the fact that proximity is measured on a relative scale (stories S3, S7), or the simplicity of causes (stories S5, S6). The model also provides quantitative laws, e.g. for the wheel of fortune near miss.

We had two negative results in the experiment (story 5, choice 2 and story 9, choice 1). Note, however, that both consist in qualitative choices, which are more prone to complex interpretations by participants. The failure in S5 is likely to result from the bad design of the story; the failure in story 9 remains a mystery (perhaps the association due to word 'retired' being used twice is sufficient in rapid readers to make the rescuers seem simple).

In its current state, this theory of luck is not as simple as it should be. There are still some conceptual connections to be done that will make the link between equations and the processing of emotional intensities more transparent. The present account is meant as an attempt to depart from mere lists of factors and to outline an integrated model of the human ability to perceive luck in events.

The research, initiated in the recent years, on the cognitive role of descriptive complexity has already produced valuable results. The model presented in this paper is meant as a contribution to this enterprise. The sensitivity to complexity differences, which is central to ST, seems to be a general law, which applies across modalities and at all levels of abstraction. Its importance in the processing of some emotions that are involved in decision processes, such as the feeling of being (un)lucky after the occurrence of an event (Loomes & Sugden 1982), should encourage further investigation in this domain.